\documentclass[12pt]{article}
\pdfoutput=1
\usepackage{graphicx}
\usepackage{epsfig}
\usepackage{epstopdf}
\DeclareGraphicsExtensions{.pdf,.eps,.png,.jpg,.mps}
\usepackage{caption}
\usepackage{float}
\usepackage{color}

\usepackage{url}
\usepackage[implicit=false]{hyperref}
\usepackage{cite}

\def\lsim{\mathrel{\rlap{\lower3pt\hbox{\hskip0pt$\sim$}}
     \raise1pt\hbox{$<$}}}         
\def\gsim{\mathrel{\rlap{\lower4pt\hbox{\hskip1pt$\sim$}}
     \raise1pt\hbox{$>$}}}         

\usepackage{amsmath}
\usepackage{amsfonts}

\begin{document}
\begin{titlepage}

\centerline{\Large \bf Non-Stationary Saturation of Inhomogeneously Broadened }
\centerline{\Large \bf Spin Systems with Effective Cross-Relaxation}
\medskip

\centerline{Zura Kakushadze$^\S$$^\dag$\footnote{\, Zura Kakushadze, Ph.D., is the President of Quantigic$^\circledR$ Solutions LLC,
and a Full Professor at Free University of Tbilisi. Email: \href{mailto:zura@quantigic.com}{zura@quantigic.com}}}
\bigskip

\centerline{\em $^\S$ Quantigic$^\circledR$ Solutions LLC}
\centerline{\em 680 E Main St \#543, Stamford, CT 06901\,\,\footnote{\, DISCLAIMER: This address is used by the corresponding author for no
purpose other than to indicate his professional affiliation as is customary in
publications. In particular, the contents of this paper
are not intended as an investment, legal, tax or any other such advice,
and in no way represent views of Quantigic$^\circledR$ Solutions LLC,
the website \url{www.quantigic.com} or any of their other affiliates.
}}
\centerline{\em $^\dag$ Free University of Tbilisi, Business School \& School of Physics}
\centerline{\em 240, David Agmashenebeli Alley, Tbilisi, 0159, Georgia}
\medskip

\centerline{(May 31, 1990; in LaTeX form: January 5, 2020)\footnote{\, This note in Russian was published in 1990 in \cite{Bulletin}. I worked on this project while still in high school. For reasons outside my control, it was a few years before it was submitted to the journal. This English translation closely follows the original Russian version, with minor changes such as equation formatting and some additional references.
}}

\bigskip
\medskip

\begin{abstract}
{}Non-stationary saturation of inhomogeneously broadened EPR lines is investigated under effective cross-relaxation. The generalized kinetic equations are solved and the complex susceptibility is calculated. All quantities change in time exponentially.
\end{abstract}
\medskip

\end{titlepage}

\newpage
{}Saturation of inhomogeneously broadened EPR (electron paramagnetic resonance) lines \cite{Portis} has been studied in detail in the stationary case. It is of substantial theoretical and applied interest to investigate the temporal attainment of the stationary state in such systems, which are described by the macroscopic parameters $\beta(\omega^\prime, t)$ and $\beta_d(t)$, the inverse temperatures of the spin packet (SP) with the frequency $\omega^\prime$ and the dipole reservoir (DR), respectively \cite{BZK1}.\footnote{\, For additional related literature, see, e.g., \cite{BZK2}, \cite{BBZ}, \cite{BZK3}, and references therein.} According to \cite{BZK1}, \cite{Thesis}, the system of the generalized kinetic equations has the following form:
\begin{eqnarray}
 &&\partial \beta(\omega^\prime, t) + {{\beta(\omega^\prime, t) - \beta_L}\over T_{SL}} + \pi\omega_1^2\varphi(\omega^\prime - \Omega)\left[\beta(\omega^\prime, t) + {{\Omega - \omega^\prime} \over \omega^\prime} \beta_d(t)\right]-\nonumber\\
 &&~~~~~~~ -{1 \over \omega^\prime}\int d\omega^{\prime\prime}~g(\omega^{\prime\prime} - \omega_0)~W_{CR}(\omega^{\prime\prime} - \omega^\prime)\times \nonumber\\
 &&~~~~~~~\times \left[\omega^{\prime\prime} \beta(\omega^{\prime\prime}, t) - \omega^\prime \beta(\omega^\prime, t) + (\omega^\prime - \omega^{\prime\prime})\beta_d(t)\right] = 0\label{Eq1}\\
 &&\partial\beta_d(t) + {{\beta_d(t) - \beta_L}\over T_{DL}} + {1 \over \omega_d^2}\int d\omega^{\prime}~g(\omega^{\prime} - \omega_0)~\omega^{\prime}~(\omega^{\prime} - \Omega)\times\nonumber\\
 &&~~~~~~~\times \left[\partial \beta(\omega^\prime, t) + {{\beta(\omega^\prime, t) - \beta_L}\over T_{SL}}\right] = 0\label{Eq2}
\end{eqnarray}
Here: $\partial$ denotes the time derivative; $\beta_L = \mbox{const}$ is the inverse temperature of the lattice; $\omega_1$ and $\Omega$ are the semi-amplitude and the frequency of the UHF (ultrahigh frequency) field; $\omega_0$ is the Zeeman frequency of the external constant magnetic field; $T_{SL}$ and $T_{DL}$ are the SP and DR spin-lattice relaxation times, respectively; $\omega_d$ is the DR energy ``quantum"; $W_{CR}(\omega^{\prime\prime} - \omega^\prime)$ is the probability of cross-relaxation (CR); $\varphi(x)$ and $g(x)$ are the homogeneous and inhomogeneous line forms, respectively.

{}In this note we study the non-stationary saturation of inhomogeneously broadened lines under effective cross-relaxation. In \cite{Thesis} the stationary limit of this case was studied, albeit the criterium for the applicability of such an approximation was not determined.

{}Let us define
\begin{eqnarray}
 &&\Delta_{CR}^2 = {{\int dx~W_{CR}(x)~x^2} \over {\int dx~W_{CR}(x)}}\\
 &&\Delta^{*2} = \int dx~g(x)~x^2
\end{eqnarray}
If the following condition
\begin{equation}
 \Delta_{CR} \gg \Delta^*\label{DeltaCR}
\end{equation}
holds, then, within the inhomogeneity of the line, the function $W_{CR}(\omega^{\prime\prime} - \omega^\prime)$ is essentially constant and in (\ref{Eq1}) we can replace it with $W_{CR}(0)$, which gives the following simplified equation:\\
\begin{eqnarray}
 &&\partial \beta(\omega^\prime, t) + {{\beta(\omega^\prime, t) - \beta_L}\over T_{SL}} + \pi\omega_1^2\varphi(\omega^\prime - \Omega)\left[\beta(\omega^\prime, t) + {{\Omega - \omega^\prime} \over \omega^\prime} \beta_d(t)\right]-\nonumber\\
 &&~~~~~~~ -{W_{CR}(0) \over \omega^\prime}\int d\omega^{\prime\prime} ~g(\omega^{\prime\prime} - \omega_0)\times \nonumber\\
 &&~~~~~~~\times \left[\omega^{\prime\prime} \beta(\omega^{\prime\prime}, t) - \omega^\prime \beta(\omega^\prime, t) + (\omega^\prime - \omega^{\prime\prime})\beta_d(t)\right] = 0\label{Eq1.simplified}
\end{eqnarray}
The system of equations (\ref{Eq1.simplified}) and (\ref{Eq2}) can be integrated in quadratures; however, the solution is very bulky. Therefore, let us assume that
\begin{equation}
 W_{CR}(0) \gg  T_{SL}^{-1} + \pi\omega_1^2\varphi(x)\label{WCR}
\end{equation}
That is, the CR is much more intensive than the saturation and the spin-lattice relaxation. The conditions (\ref{DeltaCR}) and (\ref{WCR}) define the criterium for effective CR. Furthermore, let us focus on the timescales
\begin{equation}\label{tCR}
 t \gg W^{-1}_{CR}(0)
\end{equation}
much longer than the characteristic CR time.

{}In the leading approximation in (\ref{DeltaCR}), (\ref{WCR}) and (\ref{tCR}), we have
\begin{equation}\label{beta}
 \beta(\omega^\prime, t) = \left[\omega_0\beta(\omega_0, t) + (\omega^\prime - \omega_0)\beta_d(t)\right]/\omega^\prime
\end{equation}

{}The deviation of the spin system from the equilibrium is due to the line saturation by the UHF field and can be deduced via the following exact equation:
\begin{eqnarray}
  &&\int d\omega^\prime~g(\omega^\prime - \omega_0)~\omega^\prime~\{\partial \beta(\omega^\prime, t) + [\beta(\omega^\prime, t) - \beta_L] / T_{SL} + \nonumber\\
  &&~~~~~~~ + \pi\omega_1^2\varphi(\omega^\prime - \Omega)[\beta(\omega^\prime, t) + (\Omega - \omega^\prime) \beta_d(t) / \omega^\prime]\} = 0\label{int}
\end{eqnarray}
This equation follows from (\ref{Eq1}) via multiplying the latter by $\omega^\prime g(\omega^\prime - \omega_0)$ and integrating over all $\omega^\prime$.

{}The system of equations (\ref{beta}), (\ref{int}) and (\ref{Eq2}) is straightforwardly integrable. The solution, with the equilibrium initial conditions $\beta(\omega^\prime, 0) = \beta_d(0) = \beta_L$ and taking into account that $T_{SL}/T_{DL} = 2$ or $T_{SL}/T_{DL} = 3$ and $\omega_d\ll \Delta^*$ \cite{Thesis}, has the following form:
\begin{eqnarray}
 &&[\beta_L - \beta_d(t)]/\beta_L = \pi\omega_1^2\Omega(\Omega - \omega_0)\tau G(\Omega - \omega_0)\left[1 - \exp(-t/\tau)\right]/\Delta^{*2}\label{R1}\\
 &&[\beta_L - \beta(\omega^\prime, t)]/\beta_L = \pi\omega_1^2\Omega\tau G(\Omega - \omega_0)[1 + (\Omega - \omega_0)(\omega^\prime - \omega_0)/\Delta^{*2}]\times\nonumber\\
 &&~~~~~~~\times\left[1 - \exp(-t/\tau)\right]/\omega^\prime\label{R2}
\end{eqnarray}
where
\begin{eqnarray}
 &&G(x) = \int dy~g(x + y)~\varphi(y)\\
 &&\tau^{-1} = T_{SL}^{-1} + \pi\omega_1^2 G(\Omega - \omega_0)[1 + (\Omega - \omega_0)^2/\Delta^{*2}]
\end{eqnarray}
In the stationary limit ($t\rightarrow \infty$), the solution given by (\ref{R1}) and (\ref{R2}) reduces to the result of \cite{Thesis}.

{}Usually, experimentally one measures not $\beta(\omega^\prime, t)$ and $\beta_d(t)$ but the absorption and dispersion signals $\chi^{\prime\prime}(\omega, \Omega, t)$ and $\chi^{\prime}(\omega, \Omega, t)$ (here $\omega$ is the frequency of the detecting field). The general formulas for these signals have the following form:
\begin{eqnarray}
 &&\chi^{\prime\prime}(\omega, \Omega, t) = \pi\chi_0(2\beta_L)^{-1}\int d\omega^\prime~g(\omega^\prime - \omega_0)~\varphi(\omega - \omega^\prime)\times\nonumber\\
 &&~~~~~~~\times[\omega^\prime \beta(\omega^\prime, t) + (\omega - \omega^\prime)\beta_d(t)]\label{chi2}\\
 &&\chi^{\prime}(\omega, \Omega, t) = {1\over\pi} \int d\xi~\chi^{\prime\prime}(\xi, \Omega, t)~(\xi - \omega)^{-1}\label{chi1}
\end{eqnarray}
where $\chi_0$ is the static susceptibility, and the integral in (\ref{chi1}) is understood in the principal value sense. The equilibrium signals are given by
\begin{eqnarray}
 &&\chi_0^{\prime\prime}(\omega) = \pi\chi_0\omega G(\omega - \omega_0)/2\\
 &&\chi_0^{\prime}(\omega) = {1\over\pi} \int d\xi~\chi_0^{\prime\prime}(\xi)~(\xi - \omega)^{-1}
\end{eqnarray}
Plugging (\ref{R1}) and (\ref{R2}) into (\ref{chi2}) and (\ref{chi1}), we get
\begin{eqnarray}
 &&\chi^{\prime\prime}(\omega, \Omega, t) = \chi_0^{\prime\prime}(\omega)\beta(\omega, t)/\beta_L\\
 &&\chi^{\prime}(\omega, \Omega, t) = \chi_0^{\prime}(\omega)\beta_d(t)/\beta_L - \pi\omega_1^2\Omega\tau G(\Omega - \omega_0)\times \nonumber\\
 &&~~~~~~~\times [1 - \omega_0(\Omega - \omega_0)/\Delta^{*2}]\left[1 - \exp(-t/\tau)\right][\chi_0^{\prime}(\omega) -
 \chi_0/2] / \omega
\end{eqnarray}

{}Let us introduce the following integral observable:
\begin{equation}
 S(\Omega, t) = \int d\omega~ \chi(\omega, \Omega, t)
\end{equation}
where ($i^2=-1$)
\begin{equation}
 \chi(\omega, \Omega, t) = \chi^{\prime}(\omega, \Omega, t) - i~\chi^{\prime\prime}(\omega, \Omega, t)
\end{equation}
is the complex susceptibility. Using the dispersion relations for $\chi(\omega, \Omega, t)$, one can show the following general result:
\begin{equation}
 S(\Omega, t) = i\pi \lim_{\omega\rightarrow\infty} \omega~\chi(\omega, \Omega, t)
\end{equation}
That is, $S(\Omega, t)$ is determined by the asymptotic of the complex susceptibility at infinity, which is a direct consequence of the causality principle.

{}Finally, let us mention that all quantities change in time exponentially, and, furthermore, substantial deviations from the equilibrium arise at timescales of order $\tau\gg 1/W_{CR}(0)$; therefore, the condition (\ref{tCR}) is not restrictive.

\end{document}